\documentclass[a4paper]{jpconf}

\usepackage{graphicx}

\begin{document}

\title{The hypothesis of the dust origin of the Broad Line Region in Active Galactic Nuclei}
\author{Bozena Czerny, Krzysztof Hryniewicz}
\address{Copernicus Astronomical Center, Bartycka 18, 00-716 Warsaw, Poland}
\ead{bcz@camk.edu.pl}

\begin{abstract}
Strong broad emission lines are the most important signatures of active galactic nuclei. These lines allowed to discover 
the cosmological nature of quasars, and at present these lines allow for convenient method of weighting the black holes 
residing in their nuclei. However, a question remains why such strong lines form there in the first place. Specifically, 
in the case of Low Ionization Lines, there must be
a mechanism which leads to an efficient rise of the material from the surface of the accretion disk surrounding a black 
hole but at the same time should not give a strong signature of the systematic outflow, as the Balmer lines are not 
significantly shifted with respect to the Narrow Line Region. We determine the effective temperature of the accretion 
disk underlying the H$\beta$ line at the basis of the time delay measured from reverberation and the simple Shakura-Sunyaev
theory of accretion disks. We obtain that this temperature is universal, and equal  $995 \pm 74$ K, independently from the 
black hole mass and accretion rate of the source. This result suggests to us that the dust formation in the disk 
atmosphere is responsible for the strong rise of the material. However, as the material gains height above the disk it
becomes irradiated, the dust evaporates, the radiation pressure force suddenly drops and the material fall back again
at the disk. Therefore, a failed wind forms. In the simple version of the model the disk irradiation is neglected, but
in the present paper we also discuss this irradiation and we use the observed variation of the Broad Line Region in 
NGC 5548 to constrain the character of this non-local non-stationary phenomenon. The current instruments cannot resolve
the Broad Line Region but future instrumentation may allow to test the model directly.  
\end{abstract}

\section{Introduction}

Broad emission lines in quasars and in other types of active galaxies have lead to the identification of those objects
as distant sources with strong non-stellar nuclear emission coming from their nuclei due to accretion onto black holes. 
Both Carl Seyfert (1943) \cite{seyfert1943} and Maarten Schmidt (1963) \cite{schmidt1963} in their seminal papers were just studying emission lines.
Since that time there was a tremendous progress in our understanding of the nature of active galaxies and in our knowledge
of the Broad Line Region (BLR) properties. Emission line measurements are also nowadays a powerful tool to measure
the mass of the central black hole (Kaspi et al. \cite{kaspi2000}; Peterson et al. \cite{peterson2004}) and these measurements done for large samples of 
quasars (e.g. Vestergaard et al. \cite{vestergaard2008}) give the insight into the evolution of galaxies.  But one question still remains to be 
answered: why this region actually forms there.

The properties of the BLR contain a hidden hint. The argument goes through three independent lines of reasoning.

First point: The lines cover a considerable fraction (0.1 - 0.3) of an AGN sky from 
the point of view of the nucleus, as estimated from the line luminosity (ref.~\cite{goad1998}). At the same time they are almost 
never along the line of sight to the nucleus
although NAL are seen in a significant fraction of objects (BAL are rare; 20\% of normal quasars~\cite{knigge2008}) and both NAL and BAL form at larger 
distances from the nucleus. There are rare reports of the transient absorption phenomenon seen in X-ray data which are likely to be due to BLR clouds~\cite{risaliti2011}. These observations indicate that the BLR is not spherical but 
flat and located close to the equatorial plane; in side-viewed objects this region is hidden from us by the 
dusty-molecular torus and in top-viewed objects the line of sights does not cross BLR. Kinetic arguments also
suggest such a flattening (e.g. Nikolajuk~\cite{nikolajuk2005}, Collin~\cite{collin2006}).

Second point: the line widths are from 1000 to 10 000 km s$^{-1}$, consistent with the Keplerian disk-like motion. 
However, lines rarely show two-component profile expected from a flattened disk-like configuration. There are
rare objects showing clearly double-peaked lines but these are mostly radio-galaxies (Eracleous \& Halpern~\cite{eracleous2003}), and some objects
hint for the disk-like profiles in the variable part of their line profiles. The majority of quasars have a single 
peak. Murray \& Chiang have shown that even a disk-like configuration can lead to single profiles but it requires 
special wind velocity profile~\cite{murray1997}. And no systematic shift (outflow velocity signature) with respect to the Narrow 
Line Region (NLR)  is seen in Balmer lines.

Third point: The careful studies of the luminosity and the line ratios, particularly in respects to the Low Ionization 
Lines like Balmer lines and Mg II (Collin-Souffrin~\cite{collin1988}), suggest that additional collisional heating is needed in 
addition to pure radiative heating estimated from the shape of the continuum emission from the central region of an active
galaxy.

All three problems point towards the BLR as flat configuration roughly in Keplerian motion but with additional turbulence
with the turbulence velocities of order of a thousand km s$^{-1}$. Such velocities, although only a fraction of the
Keplerian speed would be 100 times higher than the sound speed, and the development of a such highly supersonic 
turbulence is not easily explained.

In our paper we offer the plausible mechanism which gives strict quantitative predictions where the LIL part of the BLR
would form. Future high resolution observations will possibly resolve the BLR region and offer direct test of the 
proposed mechanism.

\section{Disk temperature underlying the BLR}

We determine the effective temperature of the accretion disk atmosphere underlying the BLR using the 
observational determination of the BLR radius from the reverberation method and combine it with the Shakura-Sunyaev
theory of accretion disks~\cite{shakura1973}. 

The scaling of the BLR with luminosity discovered by Kaspi et al.~\cite{kaspi2000} has been studied by 
several authors (e.g. Peterson et al.~\cite{peterson2004}). Here we use
the result of Bentz et al.~\cite{bentz2009}, where the starlight contamination was carefully removed. 
The power-law index in the scaling between the BLR size and the monochromatic luminosity in 
their analysis is consistent with 0.5 within the errors. We thus fixed this slope at 0.5, 
and refitted their data again, obtaining the relation    
\begin{equation}
\log R_{BLR}[{\rm H}\beta] = 1.538 \pm 0.027 + 0.5 \log L_{44,5100},
\label{Bentz}
\end{equation}
where $R_{BLR}[{\rm H}\beta]$ is in light days and $L_{44,5100}$ is the monochromatic luminosity at 
5100 \AA~ measured
in units of $10^{44}$ erg s$^{-1}$. The value 0.027 is the error in the best-fit vertical 
normalization; the dispersion around the best fit is larger but still small at 0.21.

We assume that the disk is not strongly irradiated, and the outflow from the disk 
is not too strong, so the simple theory of the 
alpha-disk (Shakura \& Sunyaev~\cite{shakura1973}) applies. In this case, the monochromatic flux at 5100 \AA~ can be calculated
from the model, if the black hole mass and the accretion rate are known. Taking the numerical 
formula of Tripp et al.~\cite{tripp1994} (corrected for a factor-of-two error, Nikolajuk, private communication),
we derive
\begin{equation}
\log L_{44,5100} = {2 \over 3} \log (M \dot M) - 43.8820 + \log \cos i,
\label{L5100}
\end{equation}
where the black hole mass, $M$, and the accretion rate, $\dot M$, are in g and g s$^{-1}$, respectively. 
This formula allows us to calculate the BLR radius from the Eq.~\ref{Bentz}, if the
product of ($M \dot M$) is known.

For completeness, we introduce a geometrical factor representing the effect of the inclination angle, $i$ for the
measured delay (for details, see Czerny \& Hryniewicz~\cite{czerny2011}). This gives the relation between the actual disk radius, $r$, and
the measured delay of H$\beta$ line:
\begin{equation}
r = {R_{BLR}[{\rm H}\beta] \over 1 + \sin i}
\label{eq:geom} 
\end{equation}
and we fix this factor adopting the average inclination $i = 39.2^{\circ}$ after Lawrence \& Elvis~\cite{lawrence2010}.

The Shakura-Sunyaev accretion disk theory allows us to calculate the effective temperature of the disk 
at that radius as a function of the black hole mass and accretion rate
\begin{equation}
T_{eff} = \left ({3GM \dot M \over 8 \pi r^3 \sigma_B}\right )^{0.25},
\label{Teff}
\end{equation}
where $\sigma_B$ is the Stefan-Boltzmann constant. Since BLR is far from the central 
black hole, the inner boundary effect can be neglected.

When we combine Eqs.~\ref{Bentz}, \ref{L5100}, \ref{eq:geom}, and \ref{Teff}, the dependence on 
the unknown mass and accretion rate vanishes, and we obtain a single value for all the sources 
in the sample
\begin{equation}
T_{eff} = 995  \pm 74 {\rm K},
\end{equation}
where the error reflects the error in the constant in Eq.~\ref{Bentz}. We stress that the resulting
value does not depend either on the black hole mass or accretion rate. The value is universal, for all
sources.

This value is interesting since it is close to the critical temperature at which the dust
can form. The distance is smaller than the distance to the dusty torus 
$R_{dust} \sim 0.4L_{45}^{1/2}$ pc (Nenkova et al.~\cite{nenkova2008}; where $L_{45}$ stands for the bolometric 
luminosity in $10^{45}$ erg s$^{-1}$) since in our case the
temperature results only from the local disk dissipation instead of the irradiation by the 
central parts of AGN.

\begin{figure}[t]
\begin{minipage}{.46\linewidth}
\includegraphics[width=15pc]{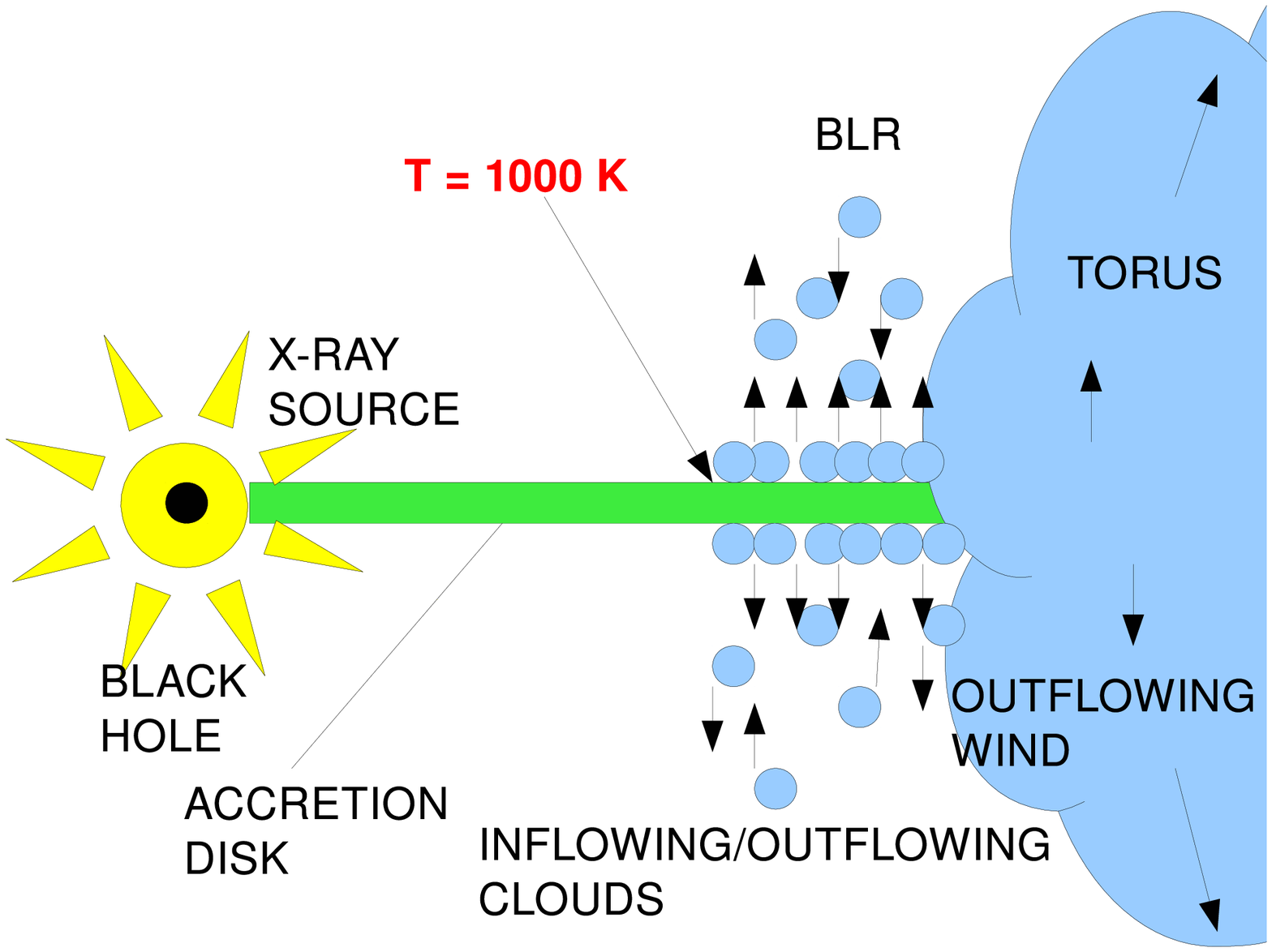}
\caption{\label{fig1}The schematic picture of the formation of BLR. In this region dusty wind rises up, the
dust evaporates due to irradiation by the central parts, gravity wins and dustless material falls back onto the disk.
Dusty torus is further from the nucleus where irradiation does not destroy the dust.}
\end{minipage} \hspace{2pc}%
\begin{minipage}{.46\linewidth}
\includegraphics[width=15pc]{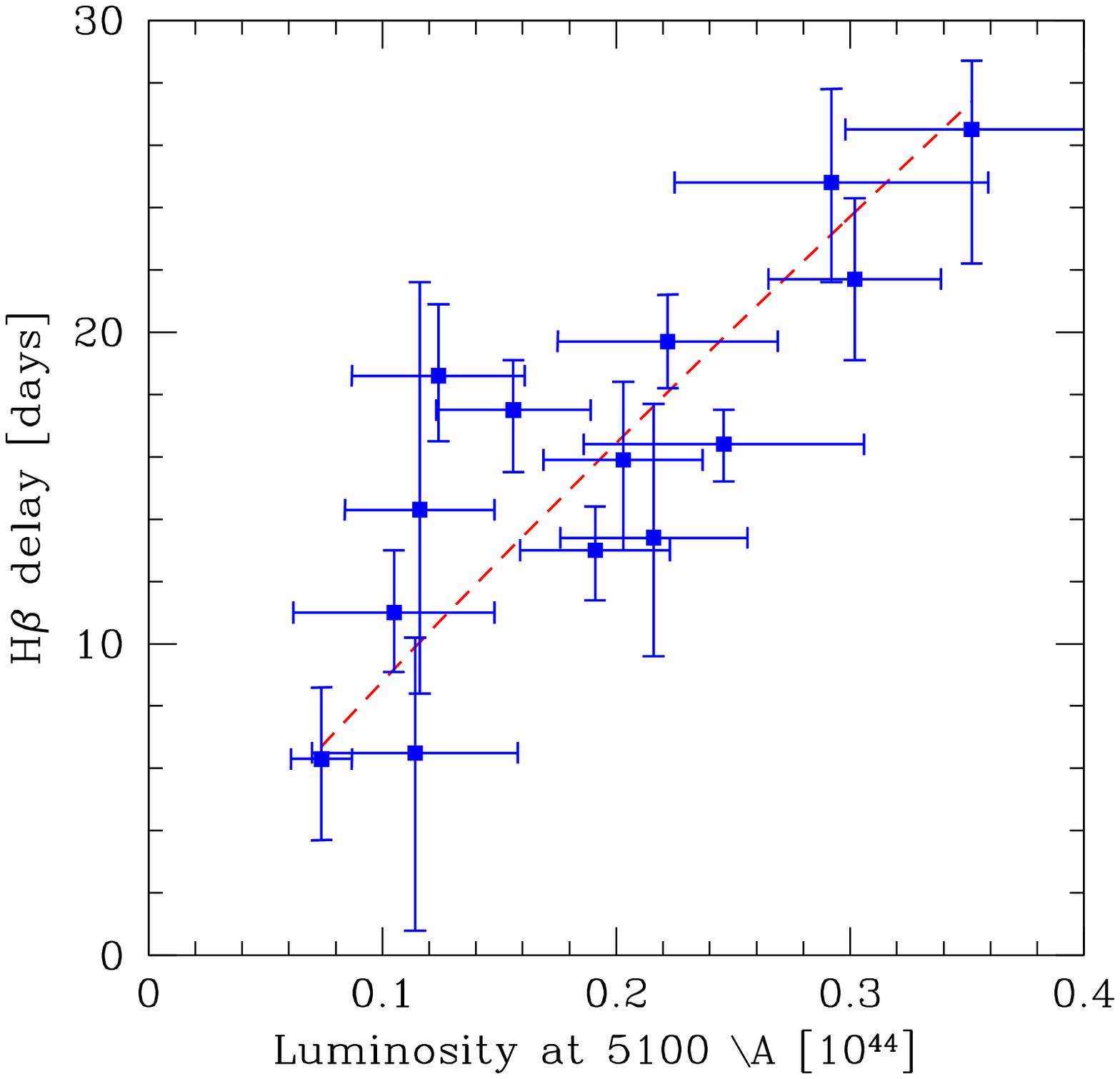}
\caption{\label{fig2}The dependence of the size of the BLR in NGC 5548 on the monochromatic luminosity: data points come from Bentz et al.~\cite{bentz2009}, and the best fit line has the slope $0.90\pm 0.22$ in the logarithmic space, instead of 0.5 characteristic for the statistical sample.}
\end{minipage} 
\end{figure}

\section{The mechanism of formation of the BLR}

This uniform effective temperature of 1000 K of the disk underlying the onset of the BLR implies that the dust present in the disk atmosphere is responsible for the phenomenon. The radiation pressure acting on dusty atmosphere is very large, as it is known for example for Wolf-Rayet stars and the strong wind develops. However, unlike in stars, the further development of the wind is modified by two factors: gravitational attraction in disk geometry rises with the height and, more importantly, the irradiation by the central parts of the accretion disk (UV) also rises very strongly due to better geometrical exposure. The dust evaporates, the rising material looses the radiation pressure support and falls back onto the disk surface. Such a failed wind can account both for the large additional turbulent velocity without any strong signature of outflow since both outflow and inflow take place with comparable velocities, and for the additional heating, as the outflowing and infalling clumps collide.

This region is separate from the region known as dusty/molecular torus. At large distances the UV irradiation is less important, the dust survives even at large height from the disk and the outflow proceeds forming an appearance of a torus. The schematic representation of the BLR formation mechanism is shown in Fig~\ref{fig1}.

\section{Discussion}

The mechanism of BLR formation proposed in this paper and in Czerny \& Hryniewicz~\cite{czerny2011} is attractive since it is universal for all sources, and the physics of dusty winds is known from stellar evolution. Although the possible presence of the dust in the BLR clouds was previously discussed by a number of authors, our idea of the key importance of the dust for BLR is new and should be further tested. The whole idea automatically explains why the size of the BLR region scales with the square root of the monochromatic luminosity, with small dispersion, since no dependence on the second parameter - for example bolometric luminosity is expected. We expect a weak dependence of the inclination angle but this also remains to be tested.

\subsection{Determination of the disk temperature in all monitored sources}

In the paper by Czerny \& Hryniewicz~\cite{czerny2011} we tested additionally the mechanism by determining the disk BLR temperature for a number of sources which were carefully monitored for many years (e.g. Peterson et al.~\cite{peterson2004}) and their luminosity was corrected for the starlight contribution at the basis of the HST data (Bentz et al.~\cite{bentz2009}). For all sources we obtain the temperature value close to 1000 K. This is also the case for NGC 5548 monitored for 14 years.

We do not see any correlation between the accretion rate and the disk effective temperature at the onset of BLR in the objects for which the reverberation monitoring was performed. Mechanism simply connecting the BLR distance with irradiation would likely contain the trend with the bolometric luminosity which (in Shakura-Sunyaev model) rises faster with the accretion rate than the monochromatic luminosity.

\subsection{The role of the disk irradiation}

In our approach to the description of the disk atmosphere underlying the BLR and to the description of the monochromatic
flux we used the simple local stationary disk model of Shakura \& Sunyaev. However, observations of AGN show that AGN 
are variable and this variability is a central point of the monitoring approach and the measurement of delays. 
It means not only that the disk is not stationary, but it implies that non-local phenomena (irradiation of the outer disk 
parts by the  inner disk) are complicated by these time delays.

The measured time delays between the continua measured at UBV bands are typically of order of a day or less in Seyfert 
galaxies (see e.g. Sergeev et al.~\cite{sergeev2005}), the major trends in optical luminosity are of order of tend-hundreds of days, 
and measured time delays of H$\beta$ line are of order of days (see compilation in Bentz et al.~\cite{bentz2009}). Both an outer and an inner disk are the sources 
of variability, as argued for example by Gaskell ~\cite{gaskell2008}, and the perturbations in the outer disk propagate 
mechanically inwards through the disk (sound waves, viscous propagation) while perturbations in the inner disk affects
the outer disk through irradiation, and the timescale is the light travel time. This complex situation can be roughly
represented with two different accretion rates/luminosities: inner accretion rate visible through the monochromatic flux
and responsible for the current strength of irradiation and the outer accretion rate (past value, likely closer to 
the average value) characteristic for the disk structure underlying the BLR.

We can now consider various possibilities depending on the importance of the irradiation on the measured value.

If the whole idea of Czerny \& Hryniewicz~\cite{czerny2011} is incorrect, and the dust formation in the accretion disk atmosphere 
has nothing to do with the formation of the BLR, then we return to the classical point of view that the onset of BLR 
requires a specific value of the ionization parameter. In this case the radius of the BLR should scale with the accretion
rate at the inner disk as
\begin{equation}
R_{BLR} \propto \dot M^{1/2}.
\end{equation}   
If we assume that the monochromatic flux is nevertheless dominated by the local dissipation, i.e. we can use 
Eq.~\ref{L5100}, the final dependence between the monochromatic luminosity and the BLR radius will be
\begin{equation}
R_{LBR} \propto L_{5100}^{0.75} M^{-1/2},
\end{equation}
i.e. the slope will be close to the older value and the presence of the additional dependence on the black hole mass
would be expected. For a given source, only the dependence on the monochromatic luminosity matters, while in a sample
of sources there is a likely bias towards positive correlation between the monochromatic luminosity and the mass, 
leading to much more shallow dependence.

Therefore, we additionally plotted the results of the monitoring for NGC 5548, taking the data points from Bentz et al.
where the starlight was subtracted. The relation is shown in Fig.~\ref{fig2}.  The slope is indeed steeper, equal to 
$0.90 \pm 0.22$. The lower limit is consistent within the error with the expectations based on the ionization parameter,
but the value itself is surprisingly high. Therefore, we considered whether the dust based scenario can explain such a 
steep slope. If we 
follow the general idea of Czerny \& Hryniewicz~\cite{czerny2011} of dust onset of the BLR but we include the irradiation of the
disk surface we also obtain the dependence of the BLR position on the accretion rate at the inner disk. Assuming that the
irradiation actually dominates we have
\begin{equation}
F_{disk} = {L_{bol} \over 4 \pi r^2}{h \over r} \sqrt {\gamma(\gamma-1)},
\label{eq:irrad}
\end{equation} 
where $h(r)$ is the disk thickness, and the radial dependence is in the form of a power law, $h(r) \propto r^{\gamma}$.

Loska et al.~\cite{loska2004} considered direct and indirect irradiation of an AGN accretion disk and calculated the disk thickness 
of the non-irradiated and irradiated disk taking into account the vertical disk structure with convection and realistic
opacities, with atomic transitions, molecules and dust. The disk thickness rises slowly with the radius at the inner part 
of the disk and then flares at larger distances, roughly as predicted by the original Shakura \& Sunyaev model. The final
rise in Loska et al.~\cite{loska2004} is roughly with $ h \propto r^{1.25}$ (see their Fig. 3). If we supplement this value of 
$\gamma$ into Eq.~\ref{eq:irrad}, assume the BLR location at a fixed temperature determined by the dust,  and combine 
it with the expression for the monochromatic flux we obtain the expected scaling
\begin{equation}
R_{BLR} \propto L_{5100}^{0.86} M^{-0.57},
\end{equation}
i.e. the trend with the black hole mass appears but the trend with the luminosity matches nicely the one seen 
for NGC 5548.

Instead, if the irradiation dominates at 5100~\AA~  the spectrum has much flatter slope ($F_{\nu} \propto \nu^{-1}$) than in the
Shakura-Sunyaev disk ($F_{\nu} \propto \nu^{1/3}$), and the monochromatic flux is proportional to the accretion rate,
$L_{5100} \propto \dot M$. In this case the combination of the irradiation dominating 5100~\AA~  and the assumption of the
specific value of the ionization parameter specified BLR gives the required relation for the statistical sample
\begin{equation}
R_{BLR} \propto L_{5100}^{0.5},
\end{equation} 
but it is inconsistent with the trends observed in NGC 5548.

\subsection{Further tests of the model}

\begin{table}
\caption{\label{tab} Observed size of the BLR in the monitored AGN.}
\begin{center}
\begin{tabular}{llllllll}
\br
name     &   redshift & $R_{BLR}$ & $R_{BLR}$ & name    &     redshift & $R_{BLR}$ & $R_{BLR}$\\
         &            &  ly     & $\mu$arc sec &    &              & ly    & $\mu$arc sec \\
\mr
Mrk335      & 0.02579 & 15.7 & 25.3  &     PG 0026+129& 0.14200 & 111.0 & 32.5 \\
PG 0052+251 & 0.15500 & 89.8 & 24.1  &     Fairall 9  & 0.04702 & 17.4  & 15.4\\
Mrk 590     & 0.02639 & 25.6 & 40.3  &     3C 120     & 0.03301 & 38.1  & 48.0\\
Ark 120     & 0.03271 & 39.7 & 50.5  &     Mrk 79     & 0.02219 & 15.2  & 28.5\\
PG 0804+761 & 0.10000 & 146.9& 61.1  &    PG 0844+349 & 0.06400 & 32.2  & 20.9\\
Mrk 110     & 0.03529 & 25.5 & 30.0  &     PG 0953+414& 0.23410 &150.1  & 26.7\\
NGC 3227    & 0.00386 & 7.8  & 84.0  &      NGC 3516  & 0.00884 & 6.7   & 31.5\\
NGC 3783    & 0.00973 & 10.2 & 43.6  &     NGC 4051   & 0.00234 & 5.8   & 103.1\\
NGC 4151    & 0.00332 & 6.6  & 82.7  &     PG 1211+143& 0.08090 & 93.8  & 48.2\\
PG 1226+023 & 0.15834 & 306.8& 80.6  &   PG 1229+204  & 0.06301 & 37.8  & 24.9\\
NGC 4593    & 0.00900 & 3.7  & 17.1  &     PG 1307+085& 0.15500 & 105.6 & 28.3\\
IC 4329A    & 0.01605 & 1.5  & 3.9   &      Mrk 279   & 0.03045 & 16.7  & 22.8\\
PG 1411+442 & 0.08960 & 124.3& 57.7  &    NGC 5548    & 0.01718 & 18.0  & 43.6\\
PG 1426+015 & 0.08647 & 95.0 & 45.7  &     Mrk 817    & 0.03146 & 21.8  & 28.8\\
PG 1613+658 & 0.12900 & 40.1 & 12.9  &    PG 1617+175 & 0.11244 & 1.5   & 26.4\\
PG 1700+518 & 0.29200 & 251.8 & 35.9 &    3C 390.3    & 0.05610 & 23.6  & 17.5\\
Mrk 509     & 0.03440 & 79.6 & 96.2  &    PG 2130+099 & 0.06298 & 22.9  & 15.1\\
NGC 7469    & 0.01632 & 4.5  & 11.5\\
\br
\end{tabular}
\end{center}
\end{table}

Better tests of the idea that the dust is the key mechanism behind the formation of the BLR require both more results
of the reverberation monitoring and better description of the actual dusty outflow.

Observationally, results from many 
years for a single source would allow to see whether the steeper relation seen for NGC 5548 is a typical pattern for 
a {\it single} source variability. Monitoring of more objects in a still broader range of masses and accretion rates
would allow to test the exact form of coupling of the BLR size to the mass and accretion rate.

Direct observational approach to the location of the BLR is not possible at present. The requested spacial resolution
is very high: the optimum source is NGC 4051 (100 $\mu$arcsec) due to its low redshift (see Table~\ref{tab}), but 
in the future even such an extreme resolution will be achieved.
Such a possibility was already envisioned by Elvis \& Karovska \cite{elvis2002}.

Theoretical studies are also needed since the actual formation of the dust in the disk atmosphere, dusty outflow,
dust/gas coupling, dust evaporation and subsequent inflow are very complex phenomena. There are papers 
addressing some of the issues 
(for example, see a recent paper of Dorodnitsyn et al. \cite{dorodnitsyn2011} about the dusty outflow forming the dusty torus), but never all of them. In addition, the dust
may acquire charge, and the presence of the large scale magnetic field may also modify the dust particle behaviour, 
(see e.g. Kovar et al.~\cite{kovar2011} for extreme examples of charges close to a black hole).

\ack
This work was supported in part by grant NN 203 380136.


\end{document}